\documentclass[prl,aps,twocolumn,superscriptaddress,showpacs,amsmath,amssymb,floatfix]{revtex4}

\usepackage{graphicx}

\newcommand{\be}{\begin{equation}}
\newcommand{\ee}{\end{equation}}

\begin{document}
\title{BCS-BEC crossover in a random external potential}
\author{G. Orso}
\affiliation{Laboratoire Physique Th\'eorique et Mod\`eles Statistiques,
Universit\'e Paris Sud, Bat. 100, 91405 Orsay Cedex, France}

\begin{abstract}
We investigate the ground state properties of a disordered superfluid Fermi gas across 
the BCS-BEC (Bose Einstein condensate) crossover. 
We show that, for weak disorder, 
 both the depletion of the condensate fraction of pairs and the normal fluid density 
exhibit a nonmonotonic behavior as a function of the interaction parameter $1/k_Fa$, reaching their
minimum value near unitarity.
We find that, moving away from the weak coupling BCS regime, Anderson's theorem ceases to apply and 
the superfluid order parameter is more and more affected by the random potential.
\end{abstract}
\maketitle

Ultracold Fermi gases near a Feshbach resonance have been the subject of intense experimental and theoretical 
investigations over the last five years \cite{rmp}. 
By tuning the inverse s-wave scattering length $1/a$ continuously from $-\infty$ to $+\infty$, the ground state
of these systems transforms from a weak-coupling BCS superfluid  into a Bose-Einstein condensate
of diatomic molecules.

An interesting and general problem is to explore the BCS-BEC crossover in the presence of a 
random external potential. 
Optical disorder can be introduced in ultracold gases in a highly controllable way.
For instance, in recent experiments with elongated  atomic BEC \cite{aspect,inguscio,germania}, speckle patterns,
created by a laser beam passing through a diffusive plate, were shown to suppress the 1D expansion of the gas.
A characteristic feature of these random potentials is that they are slowly varying in space,
with a typical length scale of the order of a few $\mu$m. 
In order to produce a real white noise disorder, which is ubiquitous in condensed matter theory, 
it was recently suggested \cite{castin} to  (randomly) trap atoms of a different species at the nodes of a tight 3D optical
lattice. If the optical polarizability of the fermions is small, the latter will not experience the periodic potential
but only the interaction with the impurity atoms.

To date, the effects of impurities on a superfluid Fermi gas near unitarity $(1/a=0)$ are
unknown, even for weak disorder.
For  s-wave BCS superconductors, 
Anderson's theorem \cite{anderson} states that the (disorder-averaged) order parameter 
 is unaffected by nonmagnetic impurities, as long as localization effects can be
neglected.  The superfluid density, measuring the dynamical response of the system to a superfluid
flow, is instead  reduced by impurities \cite{abrikosov}.
For Bose  superfluids, the depletions of the condensate fraction and of the superfluid density induced by a weak disorder
 have been calculated in Refs \cite{huang,giorgini94}. 

In this Letter we provide the first theoretical study of a disordered superfluid Fermi gas
 across the BCS-BEC crossover, at zero temperature. We show that
the depletion of the condensate fraction of pairs and the normal fluid density, induced by a weak disorder,
exhibit nonmonotonic behavior as a function of the interaction strength, with a pronounced minimum near 
unitarity.
These results are therefore consistent with the general expectation that superfluidity is more
robust against suppressing mechanisms in the crossover regime. Nonmonotonic behaviors
across the resonance have recently been observed \cite{ketterle} in critical velocities for superfluid flow,
in agreement with earlier theoretical predictions \cite{sensarma, strinati-Josephson}.

Our work is based on the Nozieres and Schmitt-Rink (NSR) theory \cite{nozieres} 
of superconducting fluctuations, extended to the broken symmetry state \cite{randeria}.
In particular, we will make use of techniques developed in Ref.\cite{griffin}, where
the NSR theory has been applied  to calculate the 
condensate fraction and the normal fluid density of {\sl clean} Fermi gases across the BCS-BEC crossover.

In the following we assume  that the random potential $V(\mathbf r)$
originates from the scattering of fermions against impurity atoms
and is given by
$V(\mathbf r)=\sum_{j} g_d \delta(\mathbf r-\mathbf R_j)$,
where $g_d$ is the {\sl fermion-impurity} coupling constant and $\mathbf R_j$ are the static positions of the impurities.
The corresponding correlation function takes the white noise form $\langle V(-q)V(q)\rangle=\beta \delta_{i\omega_m,0}\kappa$, where $q=(\mathbf q,i\omega_m)$ and $\kappa=n_i g_d^2$, $n_i$ being 
the concentration of impurities.

The two-component Fermi gas is described by 
the imaginary time action
$S=\int_0^\beta d\tau \int d\mathbf r \{\bar \psi_\sigma \partial_\tau \psi_\sigma +H\}$, where $\beta$ in the inverse temperature and
\begin{equation}\label{ham}
H=\bar \psi_\sigma \left(K +V \right )\psi_\sigma-
g \bar \psi_{\uparrow}^\dagger \bar \psi_{\downarrow}^\dagger  \psi_{\downarrow}\psi_{\uparrow}.
\end{equation}
Here $K=-\nabla^2/2m-\mu$, $\mu$ is the chemical potential, $g$ is  the fermion-fermion
interaction  and we set $\hbar=\textrm{Volume}=k_B=1$.

Following Ref.\cite{griffin}, we apply the Hubbard-Stratonovich transformation to  the Hamiltonian (\ref{ham}), 
by introducing a new bosonic field $\Delta(x)$ which couples to
$\psi_{\uparrow} \psi_{\downarrow}$. Integrating out the fermionic fields, 
the partition function of the system takes the path integral form $Z=\int D[\Delta, \bar \Delta] e^{-S_\textrm{ eff}}$, where 
\begin{equation}\label{seff}
S_\textrm{ eff}= \int_0^\beta d\tau \int d\mathbf r \left\{ \frac{|\Delta(x)|^2}{g} -\frac{1}{\beta} \textrm{ Tr} \ln[ -\beta \mathbf G^{-1}(x)]\right\}
\end{equation}
is the  effective action  written in terms of the inverse single particle Green's function
\begin{equation}\label{green}
\mathbf G^{-1}(x)=-\partial_\tau-(K+V(\mathbf r) ) \tau_3+\Delta(x)\tau^+ + \overline{\Delta}(x)\tau^-.
\end{equation}
Here $x=(\mathbf r,\tau)$, $\tau_i$ are the Pauli matrices and $\tau^{\pm}=(\tau_1\pm i \tau_2)/2$.

To procede further, we assume  that the main contribution to the partition function comes from small 
fluctuations $\delta \Delta(x)=\Delta(x)-\Delta$ 
around the {\sl uniform} BCS pairing field $\Delta$. Hence, we expand
the effective action  (\ref{seff})  up to quadratic order in the bosonic fields $\delta\Delta$ and in the 
random potential  $V$.
To this purpose, the Green's  function (\ref{green}) is conveniently written as $\mathbf G^{-1}=\mathbf G_0^{-1}+\mathbf \Sigma$, where 
$\mathbf G_0^{-1}=-\partial_\tau-K \tau_3+\Delta \tau_1$ 
is  the inverse BCS  Green's function in the absence of disorder, and
$\mathbf \Sigma=-V(\mathbf r) \tau_3+\delta \Delta(x)\tau^+ + \overline{\delta\Delta}(x)\tau^-$.

Averaging over the positions of the impurity atoms, we obtain $S_\textrm{ eff}=S_F+A (\delta \Delta(0)+\overline{\delta\Delta}(0))+S_B$, where $S_F=\beta \Delta^2/g-\beta^{-1}\sum_{k} \textrm{Tr} \ln[ -\mathbf G_0^{-1}(k)+V(0)\tau_3]+\beta \Omega_F^d$ corresponds to the 
mean field action. Here $\mathbf G_0^{-1}(k)=i \omega_n-\xi_\mathbf k \tau_3+\Delta \tau_1$  is written in momentum representation,  with $\xi_\mathbf k=|\mathbf k|^2/2m-\mu$ and
\begin{equation}\label{sF}
\Omega_F^d= \frac{1}{2\beta}\sum_{k q} 
\textrm{Tr} [\mathbf G_0 \tau_3 \mathbf G_0^\prime\tau_3] \langle V(q) V(-q) \rangle,
\end{equation}
where $\mathbf G_0=\mathbf G_0(k)$ and $\mathbf  G_0^\prime=\mathbf G_0(k+q)$.
Furthermore, $A=\Delta (1/g-\sum_\mathbf k \tanh(\beta E_\mathbf k /2)/2E_\mathbf k)$,
where $E_\mathbf k=(\xi_\mathbf k^2+\Delta^2)^{1/2}$ is the energy spectrum of  
single particle excitations, and
\begin{equation}\label{sB}
 S_B=\frac{1}{2}\sum_q \left[ \eta^\dagger \mathbf M \eta+ V(-q)W^\dagger \eta+V(q)W \eta^\dagger \right]
\end{equation}
corresponds to the  gaussian action  for the bosonic fluctuations. Here  $\eta^\dagger =[\delta \bar \Delta(q),\delta \Delta(-q)]$
and  $\mathbf M$ is a 2x2 symmetric matrix 
whose elements are given by 
\begin{equation}\label{M}
 M_{11}(q)=\frac{1}{g}+\sum_k G_{0,22} G^\prime_{0,11},\;\;
M_{12}(q)=\sum_k G_{0,12} G^\prime_{0,12},
\end{equation}
and  $M_{22}(q)=M_{11}(-q)$ (explicit expressions can be found in Ref.\cite{griffin}).
In Eq.(\ref{sB}), the doublet
\begin{equation}\label{Wu}
W(q)=\begin{pmatrix}
         \sum_k  G_{0,12}  G^\prime_{0,11}-G_{0,22} G^\prime_{0,12}\\
          \sum_k  G_{0,11}  G^\prime_{0,12}-G_{0,12} G^\prime_{0,22}
\end{pmatrix},
\end{equation}
couples disorder to the bosonic fluctuations
and induces processes in which pairs are scattered in or out of the condensate. 
At zero temperature, 
 $W_1=W_2= \sum_\mathbf k \Delta (\xi_\mathbf k+\xi_{\mathbf k+\mathbf q})/2E_\mathbf k E_{\mathbf k+\mathbf q}(E_\mathbf k+E_{\mathbf k+\mathbf q})$, for $q=(\mathbf q,0)$.

By assumption, the mean field action $S_F$ is an extremum of $S_\textrm{eff}$, 
hence the coefficient  $A$ must be set equal to zero $(A=0)$. This yields the usual BCS gap 
equation $1/g=\sum_\mathbf k \tanh(\beta E_\mathbf k /2)/2E_\mathbf k $, 
{\sl as in the absence of impurities}.
This equation is ultraviolet divergent and is cured \cite{randeria} by replacing the bare coupling constant $g$ with the scattering length $a$ according to $1/g=-m/4\pi a+\sum_\mathbf k 1/2\epsilon_\mathbf k$, with 
$\epsilon_\mathbf k=\mathbf k^2/2m$. At zero temperature, one finds
\begin{equation}\label{gap}
-\frac{m}{4\pi a}=\sum_\mathbf k \left[\frac{1}{2E_\mathbf k}-\frac{1}{2\epsilon_\mathbf k}\right],
\end{equation}
which should be seen as a self-consistent relation between $\Delta$ and $\mu$.

In the NSR theory, the second constraint is provided by the number equation $n=-\partial \Omega/\partial \mu$,
where $\Omega=-\ln Z/\beta$ is the thermodynamic  potential. To calculate it, we substitute 
$S_\textrm{ eff}=S_F+S_B$ in the partition function $Z$
and integrate over the Bose fields.
At zero temperature, this gives $\Omega=\Omega_F+\Omega_B$, where $\Omega_F=S_F/\beta$  is the mean field 
fermionic term and 
\begin{equation}\label{omegaB}
\Omega_B=\lim_{\beta\rightarrow \infty}\frac{1}{2\beta} \sum_q \ln \textrm{det} M- \frac{\kappa}{2} \sum_{\mathbf q,\omega_m=0}
 W^\dagger {\bf M}^{-1}W
\end{equation}
gives the contribution from bosonic fluctuations.

The integration over $\mathbf q$ in Eq.(\ref{sF}) is also ultraviolet divergent, due to the zero range nature of  the 
fermion-impurity potential. This problem is solved by a 
renormalization of the related coupling constant 
$g_d\rightarrow g_d+g_d^2\sum_\mathbf q 1/\epsilon_\mathbf q$. Equation (\ref{sF}) then yields
  $\Omega_F^d=\kappa \Delta^2 m^3/4 \pi^2$, 
which is independent of $\mu$.
As a result, the number equation  takes the form
\begin{equation}\label{numbereq}
n=\sum_\mathbf k \left (1-\frac{\xi_\mathbf k}{E_\mathbf k}\right)-\frac{\partial \Omega_B}{\partial \mu}.
\end{equation}
Differently from Eq.(\ref{gap}), Eq.(\ref{numbereq}) depends 
{\sl explicitly} on disorder through the contribution (\ref{omegaB}) from bosonic fluctuations.
From Eqs  (\ref{gap}) and (\ref{numbereq}) we obtain  $\Delta$ and $\mu$ as a function of the atom density
$n=k_F^3/3\pi^2$, the scattering length $a$ and the disorder strength $\kappa$ (up to linear terms). 

In the BCS limit, corresponding to  $1/k_F a \rightarrow -\infty$, the bosonic contribution  in Eq.(\ref{numbereq}) is negligible and we find $\mu= \epsilon_F=k_F^2/2m$ while Eq.(\ref{gap}) gives $\Delta=8e^{-2}\epsilon_F\exp(-\pi/2 k_F |a|)$, as in the absence of disorder. This means that for weak attraction  the superfluid order parameter is unaffected by the disorder, in agreement with Anderson's theorem \cite{anderson}.

In the opposite strong coupling regime, corresponding to $1/k_F a \rightarrow +\infty$, the gap equation
(\ref{gap}) yields $\mu=-1/2ma^2$, which is half of the  
binding energy. To find $\Delta$, we first differentiate
Eq.(\ref{omegaB}) with respect to $\mu$ and then expand the obtained result in powers of $q$, assuming
$|\mathbf q|^2/2m,i\omega_m \ll |\mu|$. Taking into account that  $W_1(\mathbf q=0)=m^2a\Delta/4\pi$, 
from Eq.(\ref{numbereq})  we obtain
\begin{equation}\label{strong}
\frac{n}{2}=\frac{m^2\Delta^2a}{8 \pi}+\sum_\mathbf q \left[ \frac{\epsilon_\mathbf q^M + \mu_M}{2 \omega_\mathbf q}-\frac{1}{2}\right] 
+4 \kappa  \sum_\mathbf q \frac{m^2\Delta^2a/8\pi}{(\epsilon_\mathbf q^M+2\mu_M)^2},
\end{equation}
where $\epsilon_\mathbf q^M=\mathbf q^2/4m$, $\omega_\mathbf q =\sqrt{\epsilon_\mathbf q^M(\epsilon_\mathbf q^M+2\mu_M})$ is the spectrum of the bosonic excitations. In Eq.(\ref{strong}),  
$\mu_M=\Delta^2 a^2 m/2=2\pi a n/m$ is the effective chemical potential for the composite bosons
corresponding to a molecular scattering length $a_M=2a$ \cite{gora}.

Equation (\ref{strong}) has a direct physical meaning: the term $(m^2\Delta^2a/8\pi)$ 
corresponds to the fraction  of condensed molecules whereas the second term yields  the 
Bogoliubov quantum depletion due to the molecule-molecule interaction, as shown in Ref.\cite{griffin}. 
The last term in Eq.(\ref{strong}) gives the density of molecules pushed out of the condensate by the random potential.
 In particular, the factor $4$ accounts for the fact that the effective disorder potential $V_M$ seen by 
the molecule is $V_M(\mathbf r)=2V(\mathbf r)$, hence the corresponding correlation function is given by
$\langle V_M(\mathbf q)V_M(-\mathbf q)\rangle =4\kappa$.   

From Eq.(\ref{strong}) we also find $\Delta-\Delta_0=-2\kappa m/\pi a$, where $\Delta_0$ is the order parameter
in the absence of disorder, showing that Anderson's theorem {\sl does not} apply in the BEC limit. 
In particular, the correction  $\Delta-\Delta_0$ diverges for 
$a \rightarrow 0^+$.  
This  signals that in the absence of repulsive interactions, the 
{\sl uniform} Bose gas of molecules is unstable against collapse into a 
localized orbital of the disorder potential \cite{huang}.

{\sl Condensate fraction of pairs}.
This is defined as the dimensionless ratio $\alpha=2 n_c/n$, 
where $n_c=\int |G_{12}(\mathbf r,\tau=0)|^2 d\mathbf r$ 
is the density of condensed pairs corresponding to 
the normalization of the Cooper pair wavefunction.
In the absence of disorder, the condensate fraction is an increasing function of the interaction strength $1/k_F a$. It has been calculated within the NSR approach in Ref.\cite{griffin}
and compared with previous mean field \cite{parola} and Quantum Monte Carlo \cite{altrostefano} predictions.

We notice that disorder 
\begin{figure}
\begin{center}
\includegraphics[width=0.85\linewidth]{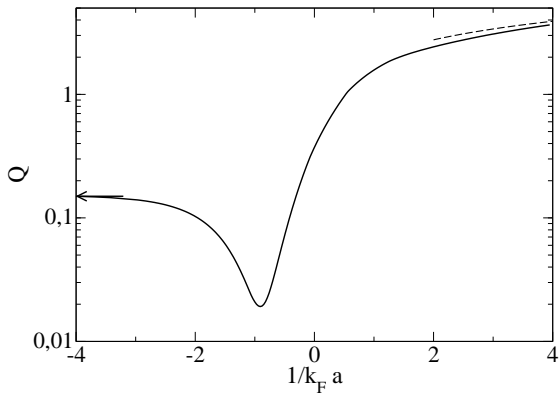}
\caption{Quantum depletion $\alpha-\alpha_0=-(\kappa m^2/k_F) Q(1/k_F a)$
of the condensate fraction  of pairs $\alpha$ induced by disorder: the function $Q(1/k_F a)$ is shown with the solid line. The asymptotic behavior at strong coupling is shown with the dashed line. Notice that $Q$ saturates
to $3/2\pi^2$ at weak coupling (arrow).}
\label{1}
\end{center}
\end{figure}
enters the formula $\mathbf G^{-1}=\mathbf G_0^{-1}+\mathbf \Sigma$
 in two different ways:  i) directly, through the ``self energy'' $\mathbf \Sigma$,
 and ii) in the mean field propagator $\mathbf G_0$, through the shifts
$\delta\mu=\mu-\mu_0,\delta\Delta=\Delta-\Delta_0$ of the variational parameters, 
as calculated from Eqs (\ref{gap}) and (\ref{numbereq}).
Hence, the change in the density of condensed pairs is given by 
\begin{equation}\label{deple}
 n_c-n_{c0}=\frac{\partial n_c^\textrm{mf}}{\partial \mu} \delta \mu+\frac{\partial n_c^\textrm{mf}}{\partial \Delta}\delta \Delta
+\sum_{\mathbf k}\frac{\Delta}{ E_\mathbf k} \int \frac{d \omega}{2\pi}\delta G^{d}_{12}
\end{equation}
where $n_c^\textrm{mf}=\sum_\mathbf k \Delta^2/4E_\mathbf k^2$ is the mean field prediction for
the condensate fraction and $\mathbf{\delta G}^{d}(k)$ corresponds to the 
disorder correction of the Green's function obtained by expanding the latter in powers of  $\mathbf \Sigma$
and retaining up to second order terms.

We have calculated the rhs of Eq.(\ref{deple}) numerically. 
We find $\alpha-\alpha_0=-(\kappa m^2/k_F)Q(1/k_F a)$,
where the crossover function $Q(1/k_F a)$ is plotted in Fig.\ref{1} with the solid line. We see that 
the depletion of the condensate fraction exhibits a pronounced minimum at $1/k_F a \approx -0.9$.

In the deep BCS limit  the leading contribution to the 
depletion (\ref{deple}) comes from $\delta G^{d}_{12}$. Neglecting  
bosonic fluctuations, we have $\delta \mathbf G^{d}(\mathbf k, i\omega)=\kappa \sum_\mathbf q  \mathbf G_0 \tau_3 \mathbf G_0^\prime \tau_3 \mathbf G_0$. 
From Eq.(\ref{deple}) we then find
 $n_c-n_{c0}=-\kappa m^2k_F^2/4\pi^4$ corresponding to $Q=3/2\pi^2$,
as shown by the arrow in Fig.\ref{1}. In the clean limit the
condensate fraction $n_{c0}=m k_F\Delta/8\pi$ 
vanishes exponentially as $1/k_F a\rightarrow -\infty$, so the effects of disorder 
are clearly important in this regime, {\sl despite the fact that the order parameter itself is not affected by impurities}.
This interesting result follows  by noticing that, in the presence of disorder, the 
zeroth-order Green's function $G_{0,12}(\mathbf r,\tau=0)$  in coordinate space is damped  
by the factor $\exp(-r/2\ell)$,
 where $\ell=\pi/\kappa m^2$  is the mean free path between collisions \cite{abrikosov}.
Hence the order parameter $\Delta=g G_{12}(\mathbf r=0,0)$ is the same as in pure superfluids, while
the condensate fraction $n_c=\int |G_{0,12}(\mathbf r,0)|^2 e^{-r/\ell}d\mathbf r\simeq \int |G_{0,12}(\mathbf r,0)|^2(1-r/\ell) d\mathbf r$ is severely suppressed by disorder. Here the condition of weak disorder
$(n_{c0}-n_c)/n_{c0} \ll 1$ yields $\kappa m k_F/\Delta \ll 1$.

In the BEC regime, the leading contribution  to the depletion (\ref{deple}) comes from the disorder induced change  
$\delta \Delta$ of the order parameter  and we 
recover $n_c-n_{c0}=-\kappa \Delta m^3/2 \pi^2 $, in agreement with Eq.(\ref{strong}) and Ref.\cite{huang}. 
The corresponding asymptotic behavior $Q= (12/\pi)^{1/2} /\sqrt{k_F a}$ is shown in Fig.\ref{1} with the dashed line. 
We see that the molecular gas is fully depleted  if the scattering length vanishes. 
Notice however that weak disorder implies
 $\kappa m^2/k_F^{3/2 }a^{1/2} \ll 1$, so one cannot 
take the limit $a\rightarrow 0^+$ for fixed disorder $\kappa$.

{\sl Normal fluid density.} This is defined as $\rho_n=n-\rho_s$, where  $\rho_s$  is the superfluid density.
The latter is related to the lowest order change $\Omega[Q_z]-\Omega[0]=\rho_s m v_s^2/2$ in the thermodynamic potential of the gas in the presence of a  supercurrent with velocity $v_s=Q_z/2m$ along, say, the z-axis.
At zero temperature and in the absence of disorder, the superfluid density coincides with the total density. Therefore $\rho_n=-\lim_{Q_z \rightarrow 0} 4m \partial^2 \Omega^d[Q_z] /\partial Q_z^2$, where $\Omega^d[Q_z]$ 
is the contribution to the thermodynamic potential coming from disorder. 

To apply this formula, we notice that
the current-carrying state is related to the equilibrium state by the gauge transformation 
$\Delta(\mathbf r)\rightarrow \Delta(\mathbf r) e^{i Q_z z}$. Accordingly,
the BCS Green's function $\mathbf G_0$ maps into $\widetilde{\mathbf G}_0$, where \cite{griffin}
\begin{equation}\label{gtilde}
\widetilde{\mathbf G}_0^{-1}(k)=\left(i \omega_n-\frac{k_z Q_z}{2m}\right)-\left(\xi_\mathbf k+\frac{Q_z^2}{8m}\right)\tau_3+\Delta \tau_1.
\end{equation}
Repeating the calculations from Eq.(\ref{sF}) to Eq.(\ref{omegaB}) with $\mathbf G_0$ replaced by
$\widetilde{\mathbf G}_0$, we obtain
\begin{equation}\label{sup}
  \Omega^d[Q_z]=\frac{\kappa}{2} \sum_{q,k}  \textrm{Tr} [\widetilde{\mathbf G}_0 \tau_3 
\widetilde{\mathbf G}_0^\prime\tau_3]-\frac{\kappa}{2} \sum_{q}
\widetilde W^\dagger \widetilde {\bf M}^{-1} \widetilde W ,
\end{equation}
where  $\widetilde{\mathbf G}_0\equiv\widetilde{\mathbf G}_0(k)$ and 
$\widetilde{\mathbf G}_0^\prime\equiv \widetilde{\mathbf G}_0(k+q)$, with $q=(\mathbf q,0)$.
In Eq.(\ref{sup}),  $\widetilde {\bf M}$ and $\widetilde W$ are given
by the expressions (\ref{M}) and (\ref{Wu}), respectively, after
replacing $\mathbf G_0, \mathbf G_0^\prime$ with $\widetilde{\mathbf G}_0, \widetilde{\mathbf G}_0^\prime$.

\begin{figure}
\begin{center}
\includegraphics[width=0.85\linewidth]{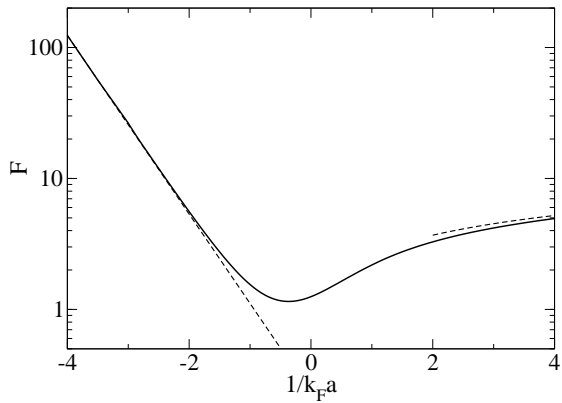}
\caption{Zero temperature normal  density  $\rho_n/n=(\kappa m^2/k_F) F(1/k_F a)$ induced by disorder: the function 
$F(1/k_F a)$ is plotted with the solid line. The asymptotic behaviors at weak and strong coupling are shown with dashed lines.}
\label{normal}
\end{center}
\end{figure}

Differentiating twice Eq.(\ref{sup}) with respect to $Q_z$ and performing the numerical integrations, we find
$\rho_n/n=(\kappa m^2/k_F)F(1/k_F a)$, where the function $F(1/k_F a)$ is plotted in 
Fig.\ref{normal} with the solid line. We see that  also the normal fluid density 
exhibits a {\sl nonmonotonic} behavior, reaching its minimum value near unitarity, at $1/k_F a \approx -0.5$. 

In the BCS regime, the leading contribution to the potential (\ref{sup}) comes from the
first term in the rhs and we find $\rho_n=\kappa m k_F^4/24 \pi^2 \Delta$, in agreement with Ref.\cite{abrikosov}.
The corresponding asymptotic behavior $F=(e^2/32)\exp(\pi/2 k_F |a|)$
 is shown in Fig.\ref{normal} with the dashed line.
In this limit the healing length of the gas is defined as $\xi_{BCS}=k_F/ m \Delta$ and we obtain 
$\rho_n/n=\pi\xi_{BCS}/8 \ell$, showing 
that the normal density is proportional to the healing length for 
$\xi_{BCS} \ll \ell$.

In the opposite BEC limit $(1/k_F a \gg 1)$, the major contribution in Eq.(\ref{sup})
comes from the bosonic fluctuations. By expanding $\widetilde W, \widetilde {\bf M}$ in
powers of $q$ as done before Eq.(\ref{strong}),
we find  $\rho_n=\kappa 4 m^3\Delta/3\pi^2$,  which is consistent with Ref.\cite{huang}.
The corresponding asymptotic behavior  $F=(8/\sqrt{3\pi})/\sqrt{k_F a}$
 is plotted in Fig.\ref{normal} with the dashed line. 

In conclusion, we have discussed the ground state properties of a disordered superfluid Fermi gas across
 the BCS-BEC crossover. We have shown that superfluidity and condensation of pairs are less affected by disorder near unitarity. Remarkably, in the BCS limit an exponentially weak disorder strongly suppresses 
the condensate fraction while the order parameter itself is insensitive to impurities, in agreement with Anderson's theorem.

An interesting direction for future work is to generalize the above theory at finite 
temperatures and to study the effect of impurities on the superfluid transition temperature $T_c$. For BCS superfluids, the 
thermodynamics of the gas (and therefore $T_c$)  is insensitive to impurities, due to Anderson's theorem. 
In contrast, for weakly interacting Bose condensates, the transition temperature is affected by disorder \cite{vinokur}.

We  acknowledge  interesting discussions with G. Falco, S. Giorgini, T. Leggett,   D. Petrov, L. Pitaevskii, N. Prokof'ev, M. Randeria, G.V. Shlyapnikov and W. Zwerger. We are also grateful to A. Griffin and E. Taylor for correspondence. 
This work is supported by the Marie Curie Fellowship under contract EDUG-038970. 

\end{document}